\documentclass[11pt,onecolumn,amssymb,nofootinbib,showpacs]{revtex4-1}
\usepackage{amsmath, amsthm, amscd, amssymb}
\usepackage{bm}
\usepackage{bbm}
\usepackage{graphicx}

\begin{document}

\title{\bf Functional Lagrange formalism for time-non-local Lagrangians}
\author{Luca Ferialdi}
\email{ferialdi@ts.infn.it}
\affiliation{Dipartimento di Fisica Teorica,
Universit\`a di Trieste, Strada Costiera 11, 34151 Trieste, Italy.
\\ Istituto Nazionale di Fisica Nucleare, Sezione di Trieste, Strada
Costiera 11, 34151 Trieste, Italy.}
\author{Angelo Bassi}
\email{bassi@ts.infn.it}
\affiliation{Dipartimento di Fisica,
Universit\`a di Trieste, Strada Costiera 11, 34151 Trieste, Italy.
 \\ Istituto Nazionale di Fisica Nucleare,
Sezione di Trieste, Strada Costiera 11, 34151 Trieste, Italy.}

\begin{abstract}
We develop a time-non-local (TNL) formalism based on variational calculus, which allows for the analysis of TNL Lagrangians. We derive the generalized Euler-Lagrange equations starting from the Hamilton's principle and, by defining a generalized momentum, we introduce the corresponding Hamiltonian formalism. We apply the formalism to second order TNL Lagrangians and we show that it reproduces standard results in the time-local limit. An example will show how the formalism works, and will provide an interesting insight on the non-standard features of TNL equations.
\end{abstract}
\pacs{03.65.Db, 11.10.Ef, 02.30.Xx}

\maketitle

\section{Introduction}
\label{sec:one}
Time-non-local (TNL) Lagrangians are generalized Lagrangians which contain integral terms involving the whole history of the system.
These Lagrangians are of interest in different fields of theoretical physics, from string theory~\cite{Eliezer:89,Aharony:00,Seiberg:99}, to regularized QFT~\cite{Pais:50,Marnelius:73}, semiclassical gravity~\cite{Seiberg:00}, non-Markovian open quantum systems~\cite{Strunz:96,Diosi:97,Diosi:98}, collapse models~\cite{Bassi:02,Adler:07,Bassi:09,Bassi2:09}, and recently light harvesting in photosynthesis process~\cite{Roden:09,Roden:11}.

A major and typical difficulty one encounters when working with TNL Lagrangians is the following. Consider for example:
\begin{equation}\label{eq:nmdiss}
L[q,\dot{q}]=\frac{m}{2}\dot{q}^2-q\int_0^t  \alpha(t,s)\big(q(s)+\dot{q}(s)\big)ds\,,
\end{equation}
where $q$ and $\dot{q}$ are respectively the position and velocity of a particle, and $\alpha(t,s)$ a memory kernel~\cite{Strunz:96,Diosi:97,Diosi:98,Bassi:09,Bassi2:09}. Suppose one wants to compute the associated Hamiltonian, and needs to transform the dependence on $(q,\dot{q})$ in a dependence on $(q,p)$ (or viceversa). The first guess would be to use the standard definition given by the Lagrange formalism: $p=\frac{dL}{d\dot{q}}$. However, one can immediately see that, since in Eq.~\eqref{eq:nmdiss} the velocity $\dot{q}$ enters the integral term, the differentiation of $L$ with respect to $\dot{q}$ does not make sense, and hence one does not know how to compute $p$.
One therefore needs to develop a new formalism which allows for the analysis of TNL Lagrangians. In doing so, one has always to make sure that, in the limit $\alpha(t,s)=\delta(t-s)$, i.e. when the Lagrangian becomes local, the standard results are recovered.

Two are the main mathematical frameworks which have been developed, to deal with TNL Lagrangians: the Ostrogradsky formalism~\cite{Ostrogradski:50,Urries:98} and the one proposed by Llosa and collaborators~\cite{Llosa:94,Gomis:01}.
The Ostrogradsky method is based on the assumption that some TNL Lagrangians can be represented as the limit of a sequence of finite higher derivative local Lagrangians. 
 In this way one can apply an extension of the standard Euler-Lagrange formalism and write the equations of motion. This is reminiscent of the well known property in probability theory, according to which a non-Markovian process can be seen as the projection of a Markovian process defined on a higher dimensional space.
It is worthwhile mentioning that the Ostrogradsky formalism cannot be applied to any TNL Action~\cite{Ostrogradski:50,Llosa:94}.

Another interesting approach is the one proposed by Llosa and collaborators~\cite{Llosa:94,Gomis:01}. 
They introduce in the Lagrangian a new parameter, in such a way that the evolution derived by this modified Lagrangian becomes local in the time variable, but non-local in the new parameter. In this way one can apply the standard Lagrange formalism with respect to the time variable, and then translate the results back to the non-local case. However, unlike the standard time-local case, the authors assume the Action to be a functional of the position only. Moreover, the formalism can be applied only to those Actions for which the functional and time derivatives commute~\cite{Llosa:94,Gomis:01}. This approach allows for some reasonable mathematical treatment, but takes away from physical intuition.
Other formalisms which have been proposed are less convincing or still debated~\cite{Woodard:00,Edelen:69}.

Aim of this Letter is to introduce a novel TNL Euler-Lagrange formalism which generalizes, in a somehow natural way, the conventional formalism, which is mathematically straightforward to handle. Moreover, such a formalism has the advantage of naturally emerging from the Hamilton's principle, and in the time-local limit automatically reproduces the standard Euler-Lagrange formalism.

\section{The formalism}\label{sec:two}
We consider a classical system described by the following TNL Lagrangian:
\begin{equation}\label{eq:ltnl}
L[q,\dot{q},t]= L_{\text{\tiny TL}}(q,\dot{q},t)+L_{\text{\tiny TNL}}[q,\dot{q},t]\,,
\end{equation}
where $L_{\text{\tiny TL}}(q,\dot{q},t)$ is the standard time local (TL) Lagrangian and $L_{\text{\tiny TNL}}[q,\dot{q},t]$ is the time non-local part, i.e. a functional of $q$ and $\dot{q}$ which accounts for the whole past history of the system:
\begin{equation}\label{eq:denovo}
L_{\text{\tiny TNL}}[q,\dot{q},t]=\int_0^t f(q,\dot{q},s)\,ds\,,
\end{equation}
with $f(q,\dot{q},s)$ a generic function. Eqs.~\eqref{eq:ltnl}-\eqref{eq:denovo} refer to a one-particle system; the generalization to many-particle systems is straightforward. Here and in the following, we denote functions with parenthesis and functionals with squared brackets. 
We require the Hamilton's principle to hold true, which means that the first variation $\delta S$ of the Action functional
\begin{equation}\label{eq:act}
S[q,\dot{q}]=\int_0^t L[q,\dot{q},s]\,ds\,,
\end{equation}
vanishes.
Note that in the standard case, that is when $L_{\text{\tiny TNL}}[q,\dot{q},t]=0$ so that the Action~(\ref{eq:ltnl}) becomes local, one has
\begin{equation}
\delta S[q,\dot{q}]=\int_0^t\left(\frac{d L_{\text{\tiny TL}}(q,\dot{q})}{dq}\delta q+\frac{d L_{\text{\tiny TL}}(q,\dot{q})}{d \dot{q}}\delta \dot{q}\right)ds\,.
\end{equation}
Integrating this expression by parts and requiring the variation to be zero, one obtains the well known Euler-Lagrange equations~\cite{Goldstein:02,Arnold:78}.
However, if $L_{\text{\tiny TNL}}[q,\dot{q},t]\neq0$ the standard Lagrange formalism cannot be directly applied, because the Action $S[q,\dot{q}]$ contains a double functional. Therefore, differentiation with respect to $q$ (or $\dot{q}$) at a fixed time does not make sense anymore. This is the reason why one needs to go beyond the standard formalism. 

A way to circumvent this difficulty is to use the identity~\cite{Gelfand:63,Greiner:96}:
\begin{equation}\label{eq:1var}
\delta S[q,\dot{q}]=\int_0^t\left(\frac{\delta S[q,\dot{q}]}{\delta q(s)}\delta q(s)+\frac{\delta S[q,\dot{q}]}{\delta \dot{q}(s)}\delta \dot{q}(s)\right)ds\,,
\end{equation}
where $\delta /\delta q(s)$ denotes the functional derivative with respect to $q(s)$. Integrating by parts Eq.~\eqref{eq:1var} one obtains:
\begin{equation}\label{eq:varpr}
\delta S[q,\dot{q}]=\int_0^t\left(\frac{\delta S[q,\dot{q}]}{\delta q(s)}-\frac{d}{ds}\frac{\delta S[q,\dot{q}]}{\delta \dot{q}(s)}\right)\delta q(s)\;ds=0\,.
\end{equation}
Comparing this result with the local one, one then naturally defines the TNL Euler-Lagrange equations as follows:
\begin{equation}\label{eq:genEL}
\frac{\delta S[q,\dot{q}]}{\delta q(s)}-\frac{d}{ds}\frac{\delta S[q,\dot{q}]}{\delta \dot{q}(s)}=0\,,\qquad\forall \,s\in[0,t]\,.
\end{equation}
Some comments are at order. This equation of motion is an integro-differential equation which depends on the whole time interval $[0,t]$. As we will see in the following example, the correct interpretation of the Eq.~\eqref{eq:genEL} will require special care. 
Secondly, one can easily check that when the Lagrangian is local, i.e. $L_{\text{\tiny TNL}}[q,\dot{q},t]=0$, Eq.~\eqref{eq:genEL} reduces to the standard Euler-Lagrange equations. 
As a third comment, our formalism looks similar to the one used for (space) non-local fields in Quantum Field Theory. However, in that case fields are non-local only in the space variables and not in the time one~\cite{Greiner:96,Weinberg:95}. The fact that in our case non-locality affects the evolution variable, drastically changes the physical situation.
We also stress that the presence of the Action $S[q,\dot{q}]$ instead of the Lagrangian in Eq.~\eqref{eq:genEL} is crucial, otherwise the formalism would be inconsistent (the functional derivative would generate Dirac deltas).

We carry on the analogy with the standard formalism, introducing a generalized momentum:
\begin{equation}\label{eq:mom}
p(s)=\frac{\delta S[q,\dot{q}]}{\delta \dot{q}(s)}\,,\qquad\forall \,s\in[0,t]\,,
\end{equation}
and defining the following generalized Hamiltonian functional:
\begin{equation}\label{eq:genH}
\mathcal{H}[q,p]=- S[q,\dot{q}]+\int_0^t p(s) \dot{q}(s)\,ds\,.
\end{equation}
The reason for introducing $\mathcal{H}[q,p]$ is that, as in the TNL Lagrangian formalism we need to express the functional equations of motion in terms of the Action instead of the Lagrangian, in the TNL Hamiltonian formalism we need to use the generalized Hamiltonian $\mathcal{H}[q,p]$ instead of the standard Hamiltonian, otherwise, infinities would appear from the functional derivatives.
We now evaluate the first variation of $\mathcal{H}[q,p]$, from which we will derive the equations of motion in the TNL Hamilton formalism. Using Eq.~\eqref{eq:genH} one can write
\begin{eqnarray}
\delta \mathcal{H}[q,p]&=&\int_0^t \left(-\frac{\delta S[q,\dot{q}]}{\delta q(s)}\delta q(s)-\frac{\delta S[q,\dot{q}]}{\delta \dot{q}(s)}\delta \dot{q}(s)+\delta p(s) \dot{q}(s)+ p(s)\delta \dot{q}(s)\right)ds\nonumber\\
&=&\int_0^t \left(-\frac{\delta S[q,\dot{q}]}{\delta q(s)}\delta q(s)+\delta p(s) \dot{q}(s)\right)ds\,,
\end{eqnarray}
where last line comes from Eq.~\eqref{eq:mom}.
Comparing this expression with the general form of the variation of $\mathcal{H}[q,p]$~\cite{Greiner:96}:
\begin{equation}
\delta \mathcal{H}[q,p]=\int_0^t\left(\frac{\delta \mathcal{H}[q,p]}{\delta q(s)}\delta q(s)+\frac{\delta \mathcal{H}[q,p]}{\delta p(s)}\delta p(s)\right)ds\,,
\end{equation}
one finds that
\begin{equation}
\label{eq:qdot}\frac{\delta \mathcal{H}[q,p]}{\delta q(s)}=-\frac{\delta S[q,\dot{q}]}{\delta q(s)}\,,\hspace{1cm}\frac{\delta \mathcal{H}[q,p]}{\delta p(s)}=\dot{q}(s)\,.
\end{equation}
This completes the TNL Hamiltonian formulation of the equations of motion. 
So far, we have developed the formalism under a rather formal point of view. We now show how this formalism applies to a specific, and very common in physics, example of TNL Lagrangian.

\section{Application of the formalism}\label{sec:three}
We consider a classical particle of mass $m$ evolving according to a general TNL second order Lagrangian, given by Eq.~\eqref{eq:ltnl} with:
\begin{eqnarray}
L_{\text{\tiny TL}}(q,\dot{q},s)&=& \frac{m}{2}\dot{q}^2(s)+A q(s)\dot{q}(s)+Bq^2(s)+C\dot{q}(s)+Dq(s)\,,\nonumber\\
L_{\text{\tiny TNL}}[q,\dot{q},s]&=&\int_0^s \alpha(s,r)\left(Eq(s)\dot{q}( r)+Fq(s)q(r)+G\dot{q}(s)q(r)+H\dot{q}(s)\dot{q}( r)\right)dr\,,
\end{eqnarray}
where the coefficients $A,\dots, H$ are constants, and $\alpha(t,s)$ is a two-variable function.
More general expressions for the Lagrangian can be considered, and the formalism still applies.
First of all, we show that such a Lagrangian, for the purposes of our formalism, can be rewritten in a simpler but equivalent way. 
The reason is that, since the crucial quantity is the Action $S$, which is the integral of the Lagrangian, then $ L_{\text{\tiny TNL}}$ can be rewritten in different ways, without changing $S$. In particular, it is convenient to integrate by parts some of the terms containing double integrals. For example, the first term of $\int_0^t L_{\text{\tiny TNL}}[q,\dot{q},s]$ can be rewritten as follows:
\begin{equation}
E\int_0^tq(s)\int_0^s \alpha(s,r)\dot{q}(r)\,dr\,ds=E\int_0^tq(s)\left(\alpha(s,s)q(s)-\alpha(s,0)q(0)-\int_0^s \frac{\partial \alpha(s,r)}{\partial r}q( r)\,dr\right)ds\,,
\end{equation}
and the first two terms on the right-hand-side can be absorbed in the term of $L_{\text{\tiny TL}}$ proportional to $q(s)$. The same happens for the third and fourth terms of $L_{\text{\tiny TNL}}[q,\dot{q},s]$.
Accordingly, without loss of generality, $L[q,\dot{q},s]$ can be rewritten as follows:
\begin{equation}
\label{eq:lex}L[q,\dot{q},s]=\frac{m}{2}\dot{q}^2(s)+\tilde{A}(s) q(s)\dot{q}(s)+\tilde{B}(s)q^2(s)+\tilde{C}(s)\dot{q}(s)+\tilde{D}(s)q(s)+\tilde{F}(s)q(s)\int_0^s \alpha(s,r)q(r)\,dr\,,
\end{equation}
where now the coefficients $\tilde{A}(s)\dots\tilde{F}(s)$ are functions of time, which collect all the terms coming from the integrations by parts described here above.
We can now use the formalism previously introduced, in particular Eq.~\eqref{eq:genEL} (or, equivalently, the generalized Hamiltonian formalism) to find the second order integro-differential equation of motion, which turns out to be:
\begin{equation}\label{eq:Fex}
m\ddot{q}(s)=-A'(s)q(s)-C'(s)+2B(s)q(s)+D(s)+\tilde{F}(s)\int_0^s\alpha(s,r) q( r)\,dr+\int_s^t\tilde{F}( r)\alpha(s,r) q( r)\,dr\,.
\end{equation}
This equation represents the generalization of the equation of motion of the standard local theory. 
The first thing one notices is that it displays a dependence on the future of $s$, not on its past, as one would naively expect by looking at the Lagrangian. The reason is that the problem was originally formulated with boundary conditions at the initial time $0$ and the final time $t$: finding the path that minimizes the Action, among all those which take the value $q_0$ at time $0$ and $\bar{q}$ at time $t$. While, in the standard time-local case, these boundary conditions can always be replaced with initial conditions on the position $q_0$ and velocity $\dot{q}_0$ at time $0$, in the TNL case this is not possible anymore, in general. This does not imply that the dynamics of the system necessarily depends on the future. It could, like in the Abraham-Lorentz equation~\cite{Jackson:99} for a charged particle, which takes back-reaction into account, and displays what is called pre-acceleration; or like in the Weehler-Feynman formulation of classical electrodynamics~\cite{Bauer:10}. But it could also not depend on the future, as it happens for non-Markovian stochastic processes, even if the dynamical equation formally does depend on the future. In this case, the equation does not correspond to the intuitive way of describing the dynamical evolution of the system. In general, there is no rule for discriminating which TNL equations can be re-written in such a way to show only a past-dependence, and which really contain a dependence on the future. From a mathematical (but also physical) point of view, this is the major difference between time-local and TNL dynamics, and why TNL dynamics are by far much more complicated. We will come back on this issue.
As a final note, it is easy to check that in the limit $\tilde{F}(s)=0$ or $\alpha(t,s)\rightarrow\delta(t-s)$ Eq.~\eqref{eq:Fex} reduces to the standard equation of motion.

In order to complete our general example, we compute the generalized Hamiltonian associated to the TNL Action of Eq.~\eqref{eq:lex}.
In particular, computing $p(s)$ with Eq.~\eqref{eq:mom}, and substituting the corresponding expression, as well as Eq.~\eqref{eq:lex}, in Eq.~\eqref{eq:genH}, one finds that:
\begin{eqnarray}\label{eq:Hex}
\mathcal{H}[q,p]&=&\int_0^t \bigg[ \frac{1}{2m}p^2(s)-\frac{A(s)}{m}q(s)p(s)+\left(\frac{A(s)}{2m}-B(s)\right)q^2(s)-\frac{C(s)}{m}p(s)\nonumber\\
&&\qquad+\left(\frac{A(s)C(s)-D(s)}{m}\right)q(s)+\frac{C^2(s)}{2m}-\tilde{F}(s)q(s)\int_0^s \alpha(s,r)q(r)\,dr\bigg]ds
\end{eqnarray}
Also in this case, one can easily check that in the time-local limit ($\alpha(s,r)\rightarrow\delta (r-s)$) this TNL Hamiltonian reduces to the standard one. Moreover, the transformation in Eq.~\eqref{eq:genH} can be reversed and, substituting in it Eq.~\eqref{eq:qdot}, one recovers $S[q,\dot{q}]$ from Eq.~\eqref{eq:Hex}.

\section{Example of a TNL system: the TNL harmonic oscillator}\label{sec:four}
We now apply the TNL formalism to a specific physical situation in order to better understand the effects of the TNL terms on the dynamics of the system. In  particular we consider a TNL harmonic oscillator, described by the following Lagrangian:
\begin{equation}\label{eq:Lho}
L[q,\dot{q}]=\frac{m}{2}\dot{q}(t)^2-\frac{k}{2}q(t)^2-\tilde{k}q(t)\int_0^t \alpha(t,s) q(s)\,ds\,,
\end{equation}
where an integral term, which accounts for the memory effects, is added to the local Lagrangian.
The equation of motion becomes:
\begin{equation}\label{eq:Fho}
m \ddot{q}(s)= -kq(s)-\tilde{k}\int_0^t\alpha(s,r) q( r)\,dr\,.
\end{equation}
Again, we can see that the equation of motion formally depends on the future up to time $t$, other than depending on the past. To show that this is not necessarily a real physical dependence on the future, let us consider an exponential memory kernel
\begin{equation}
\alpha(t,s)=\frac{\gamma}{2}e^{-\gamma |t-s|}\,,
\end{equation}
and show how the integro-differential equation can be transformed into an ordinary differential equation. (Note that by taking the limit $\gamma\rightarrow\infty$ one recovers the time local case.) With this choice for the memory kernel,
Eq.~\eqref{eq:Fho} becomes:
\begin{equation}\label{eq:hoexp}
m \ddot{q}(s)= -kq(s)-\frac{\tilde{k}\gamma}{2}\int_0^t e^{-\gamma |r-s|}q( r)\,dr\,.
\end{equation}
Note this is the same equation one obtains by applying the formalism of Llosa et al.~\cite{Llosa:94} to the Lagrangian~\eqref{eq:Lho}.
Differentiating Eq.~\eqref{eq:hoexp} with respect to $s$ one finds 
\begin{equation}\label{eq:ho3}
\dddot{q}(s)-\frac{k}{m}\dot{q}(s)-\frac{\tilde{k} \gamma^2}{2m}\int_0^s e^{-\gamma(s-r)}q( r)\,dr-\frac{\tilde{k} \gamma^2}{2m}\int_s^t e^{-\gamma(r-s)}q( r)\,dr=0\,,
\end{equation}
and differentiating it once more and using Eq.~(\ref{eq:hoexp}), one can show that $q(s)$ satisfies the following fourth order differential equation:
\begin{equation}\label{eq:hoq4}
\ddddot{q}(s)-\left(\frac{k}{m}+\gamma^2\right)\ddot{q}(s)+\frac{\gamma^2}{m}(k+\tilde{k})q(s)=0\,.
\end{equation}
The first important message is that this is an ordinary differential equation, with no explicit dependence on the future (or past). This makes the point that, quite generally, one needs special care when interpreting the meaning of the solutions of integro-differential equations.

The space of solutions of Eq.~(\ref{eq:hoq4}) is larger than that of Eq.~\eqref{eq:hoexp}. Therefore, one has to force suitable constraints on the solutions of Eq.~\eqref{eq:hoq4} in order for them to solve Eq.~\eqref{eq:hoexp} as well.
 In particular, since Eq.~\eqref{eq:hoq4} is a fourth order differential equation, four boundary conditions are needed, to univocally determine the solution. Two of them are given  by the boundary conditions of the variational problem, $q(0)=q_0$ and $q(t)=\bar{q}$. The dependence on the endpoint can be removed, like in the time local formalism, by replacing it with the initial condition on velocity $\dot{q}(0)=v_0$.
The other two boundary conditions are given by the \lq\lq consistence conditions\rq\rq~of Eq.~\eqref{eq:hoq4} with Eq.~\eqref{eq:hoexp}, i.e. by asking the solutions of Eq.~\eqref{eq:hoq4} to be solutions of Eq.~\eqref{eq:hoexp} as well.
 In order to find these two constraints, one evaluates Eqs.~\eqref{eq:hoexp}-\eqref{eq:ho3} at time $s=0$ and then at time $s=t$, and compares all terms of these equations which are equal. The result is:
 \begin{eqnarray}
 \dddot{q}(0)-\frac{k}{m}\dot{q}(0)&=& \gamma\left(\ddot{q}(0)-\frac{k}{m}q(0)\right)\,,\\
\dddot{q}(t)-\frac{k}{m}\dot{q}(t)&=& -\gamma\left(\ddot{q}(t)-\frac{k}{m}q(t)\right)\,. \label{eq:safs}
\end{eqnarray}
The first consistency condition can be considered as an extra initial condition. The second consistency condition, instead, refers to the terminal time $t$. We are not aware of any way to replace it with an additional initial condition at time 0; accordingly, our original problem with boundary conditions at $s=0$ and $s=t$ can only partly be reformulated as a problem with initial conditions at $s=0$. But once again, we stress that this does not imply that the dynamics depends on the future. It is a consequence of the fact that the original problem was formulated with boundary conditions both at time 0 and at time $t$.

The solution of Eq.~\eqref{eq:hoq4} reads:
\begin{equation}\label{eq:q4}
q(s)= a_{1,t} \sinh x_1 s + a_{2,t} \cosh x_1 s +a_{3,t} \sinh x_2 s +a_{4,t} \cosh x_2 s\,,
\end{equation}
where the subscript $t$ denotes the dependence on the endpoint, due to the condition~(\ref{eq:safs}). The roots $x_i$ and the coefficients $a_{i,t}$ are uniquely determined, and their explicit expression is given in the Appendix. We now discuss the physical behavior of the solutions.

\begin{figure}
\begin{center}
\includegraphics[width=8.8cm]{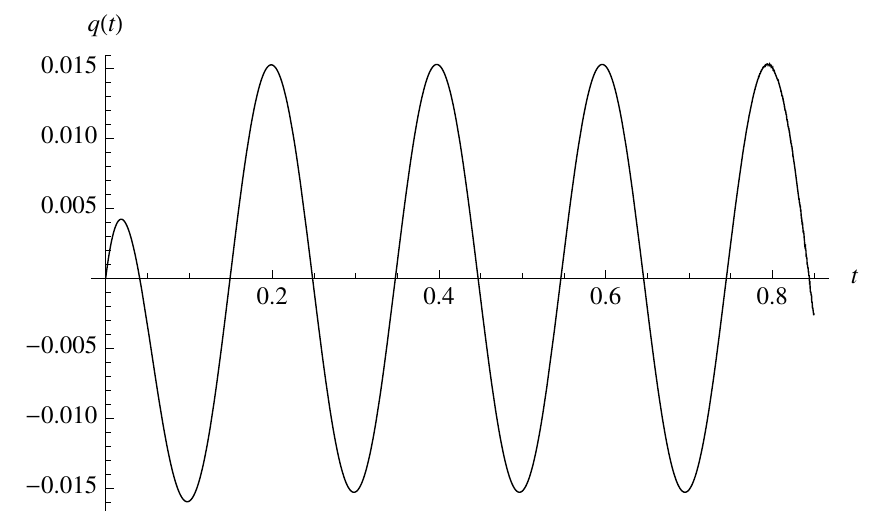}
\caption{Time evolution of $q(s)$ for a TNL harmonic oscillator of mass $m=1$, with initial values $x_0=0$, $v_0=1$. The spring constants are $k=1$, $\tilde{k}=10^6$, and the memory kernel cutoff is $\gamma=1$.} \label{fig:1}
\end{center}
\end{figure}

Figure~\ref{fig:1} shows the behavior of a TNL harmonic oscillator of mass $m=1$, with initial position $q_0=0$ and initial speed $v_0=1$. The strength of the coupling constants $k$, $\tilde{k}$, and the correlation function cutoff $\gamma$, have been chosen in such a way to highlight memory effects: $k=1$, $\tilde{k}=10^6$, and $\gamma=1$. We see that, as expected from a dynamics with memory effects, after an initial transient of few oscillations, the trajectory stabilizes and becomes harmonic. We notice that, if also the initial velocity is $v_0=0$, the oscillator stays at rest like in the time-local case.

\begin{figure}
\begin{center}
\includegraphics[width=8.8cm]{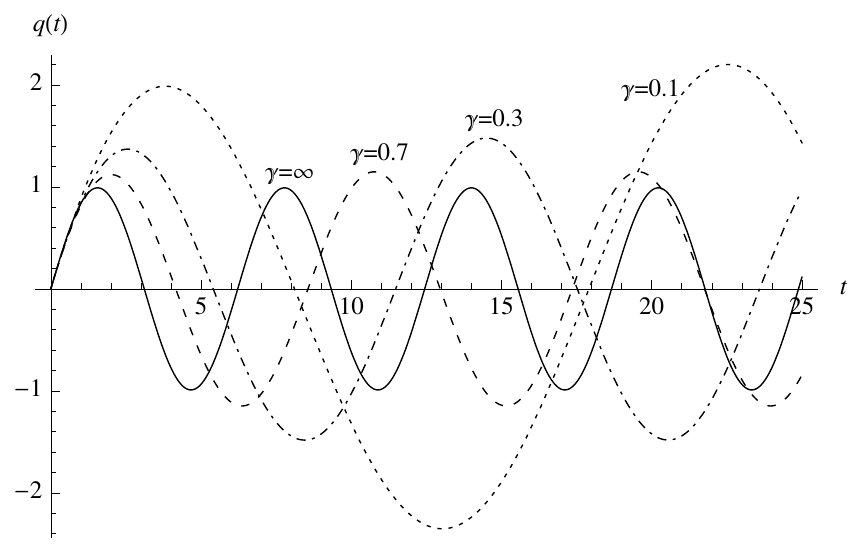}
\caption{Time evolution of $q(s)$ of an oscillator with $m=1$, for different values of cutoff $\gamma$ (dashed lines). A comparison with the standard harmonic oscillator (solid line) is given. The input data are: $x_0=0$, $v_0=1$, $k=1$, $\tilde{k}=10^6$.} \label{fig:2}
\end{center}
\end{figure}

Figure~\ref{fig:2} shows a comparison between the behavior of a standard harmonic oscillator ($m=1$, $v_0=1$, $\gamma=\infty$) and those of TNL harmonic oscillators (same $m$ and $v_0$) for different values of cutoff: $\gamma=0.1$, $0.3$, $0.7$.
As we can see, for very shorts times, the behavior of the non-local oscillator follows that of a local one. 
Moreover, as expected, the bigger $\gamma$, the closer the non-local trajectories are to the local one.
We further notice that the wavelength and the amplitude of the oscillations to which the TNL oscillator stabilizes are bigger than the corresponding local oscillator.

We can then identify three main stages in the evolution of the TNL oscillator: 

\begin{enumerate}

\item {\it Very-short time behavior}: The TNL harmonic oscillator behaves as the local one. The reason is that the elapsed time is so short that there cannot be any memory of the past dynamics yet. In fact, in this stage the contribution coming from the integral term of Eq.~\eqref{eq:hoexp} is negligible. 

\item {\it Transient phase}: In this stage, the oscillator behaves in an highly non harmonic way; this is when the TNL terms start affecting the dynamics. 
The length of this stage clearly depends on the non-local coupling constant $\tilde{k}$ and memory kernel cutoff $\gamma$. 

\item {\it Long-time behavior}: The trajectory stabilizes to that of a local harmonic oscillator. The reason is that when the elapsed time is larger than the characteristic time of the memory kernel, the integral term of Eq.~\eqref{eq:hoexp} reaches, for all practical purposes, an asymptotic stable value, and the subsequent dynamics looks \lq\lq local-like\rq\rq. The long-time wavelength of the TNL oscillator is larger than that of a local one ($\gamma=\infty$), because the effective force to which the oscillator is subject is weaker. The reason is that when $q(s)$ is, like in the present case, an oscillating function, the value of the integral of Eq.~\eqref{eq:hoexp} for any finite cutoff $\gamma$ is always smaller than the local case ($\gamma\rightarrow\infty$). 

\end{enumerate}

\noindent These features of the TNL harmonic oscillator correspond to the physical intuition one would have about a standard harmonic oscillator interacting with a non-Markovian bath.

\section{Conclusions}
We have proposed a TNL formalism which allows for a general analysis of TNL Lagrangians, and we have derived the corresponding Hamilton formalism. 
As we have discussed, to study TNL dynamics starting from their Action, one has to solve a problem with boundary conditions at time 0 and time $t$, which in general cannot be reduced to an initial-value problem (contrary to the standard time-local case, where this can always be done). This can be embarrassing in some cases, since the solution of the problem turns out to formally depend on the future. However, in important situations---like the path integral formalism---one needs only to solve a problem with boundary conditions at the initial and final time, without any need to transform in into an initial-value problem. Therefore, the TNL formalism can be applied to such a formalism without difficulties, other than technical ones related to the evaluation of the integrals.

\acknowledgements
 The authors wish to thank D. D\"urr for illuminating discussions on this subject. They acknowledge partial financial support from MIUR (PRIN 2008), INFN, COST (MP1006) and the John Templeton Foundation project \lq Quantum Physics and the Nature of Reality\rq.

\section{Appendix}

We provide the explicit expressions for the roots $x_i$ and the coefficients $a_i$ of Eq.~\eqref{eq:q4}, which determine the solution of Eq.~\eqref{eq:hoq4} for the given initial conditions. 
The roots $x_i$ are the solutions of the characteristic polynomial associated to Eq.~\eqref{eq:hoq4} and they read ($i=1,2$):
\begin{equation}
x_i=\sqrt{\frac{\gamma^2}{2}-\frac{k}{m}+(-1)^{i+1}\sqrt{\left(\frac{\gamma^2}{2}-\frac{k}{m}\right)^2+\gamma^2\frac{\tilde{k}}{m}}}\,.
\end{equation}

Since the coefficients $a_i$ have the same denominator $d$, is it useful to define four new coefficients $b_i$ as follows:
\begin{equation}
b_{i,t}=a_{i,t} d\,,\qquad i=1,\dots 4\,.
\end{equation}

The explicit expression for the denominator $d$ and the coefficients $b_i$ are:
\begin{eqnarray}
d&=&(x_1^2-x_2^2)\left[-2 x_1 x_2 \left(\frac{k}{m}+x_2^2\right) \gamma  \cosh x_2 t+\left(\frac{k}{m}+x_1^2\right) x_2 \left(2 x_1 \gamma  \cosh x_1 t+\left(x_1^2+\gamma ^2\right) \sinh x_1 t\right)\right.\nonumber\\
&&\left.-x_1 \left(\frac{k}{m}+x_2^2\right) \left(x_2^2+\gamma ^2\right) \sinh x_2 t\right]\,,\\
b_{1,t}&=&\left(\frac{k}{m}+x_2^2\right) \bigg[-x_2 \gamma  \left(v_0 \left(-\frac{k}{m}+x_1^2-2 x_2^2\right)+x_0 \left(\frac{k}{m}+x_1^2\right) \gamma \right) \cosh x_2 t\nonumber\\
&&-\left(\frac{k}{m}+x_1^2\right) x_2 (v_0-x_0 \gamma ) (\gamma  \cosh x_1t+x_1 \sinh x_1 t)\nonumber\\
&&+\left(v_0 x_2^2 \left(\frac{k}{m}+x_2^2\right)-x_0 \left(\frac{k}{m}+x_1^2\right) x_2^2 \gamma +v_0 \left(-x_1^2+x_2^2\right) \gamma ^2\right) \sinh x_2 t\bigg]\,,\\
b_{2,t}&=&x_0 \,d- b_{4,t}\,,
\end{eqnarray}
\begin{eqnarray}
b_{3,t}&=&x_1 \left(\frac{k}{m}+x_1^2\right) \bigg[\left(\frac{k}{m}+x_2^2\right) \gamma  (-v_0+x_0 \gamma ) \cosh x_2 t\nonumber\\
&&-x_1 \gamma  \left(v_0 \left(-\frac{k}{m}-2 x_1^2+x_2^2\right)+x_0 \left(\frac{k}{m}+x_2^2\right) \gamma \right) \cosh x_1 t\nonumber\\
&&-\left(-v_0 x_1^2 \left(\frac{k}{m}+x_1^2\right)+x_0 x_1^2 \left(\frac{k}{m}+x_2^2\right) \gamma +v_0 \left(-x_1^2+x_2^2\right) \gamma ^2\right) \sinh x_1 t \nonumber\\
&&-x_1 x_2 \left(\frac{k}{m}+x_1^2\right) \left(\frac{k}{m}+x_2^2\right) (v_0-x_0 \gamma ) \sinh x_2 t\bigg]\,,\\
b_{4,t}&=&\left(\frac{k}{m}+x_1^2\right) \bigg[x_1 x_2 \left(-v_0 \left(\frac{k}{m}+x_2^2\right)-x_0 \left(-\frac{k}{m}-2 x_1^2+x_2^2\right) \gamma \right) \cosh x_1 t\nonumber\\
&&+x_2 \left(-v_0 \left(\frac{k}{m}+x_2^2\right) \gamma +x_0 \left(\frac{k}{m} \gamma ^2+x_1^2 \left(x_1^2-x_2^2+\gamma ^2\right)\right)\right) \sinh x_1 t\nonumber\\
&&+x_1 \left(\frac{k}{m}+x_2^2\right) (v_0-x_0 \gamma ) (x_2 \cosh x_2 t+\gamma  \sinh x_2 t)\bigg]\,.\end{eqnarray}

\end{document}